\documentclass[english]{article}
\usepackage[T1]{fontenc}
\usepackage[latin9]{inputenc}
\usepackage{geometry}
\usepackage{xcolor}
\usepackage{pdfcolmk}
\usepackage{amsmath}
\usepackage{graphicx}
\PassOptionsToPackage{normalem}{ulem}
\usepackage{ulem}

\makeatletter

\providecolor{lyxadded}{rgb}{0,0,1}
\providecolor{lyxdeleted}{rgb}{1,0,0}
\DeclareRobustCommand{\lyxsout}[1]{\ifx\\#1\else\sout{#1}\fi}

\newcommand{\lyxaddress}[1]{
	\par {\raggedright #1
	\vspace{1.4em}
	\noindent\par}
}

\makeatother

\usepackage{babel}
\begin{document}
\title{Instanton effects vs resurgence in the $O(3)$ sigma model}
\author{Zoltán Bajnok$^{1}$, János Balog$^{1}$, Árpád Heged\H{u}s$^{1}$
and István Vona$^{1,2}$}
\maketitle

\lyxaddress{\begin{center}
~\\
$^{1}$\emph{Wigner Research Centre for Physics}\\
\emph{Konkoly-Thege Miklós u. 29-33, 1121 Budapest, Hungary}
\par\end{center}}

\lyxaddress{\begin{center}
$^{2}$\emph{Roland Eötvös University, }\\
\emph{Pázmány sétány 1/A, 1117 Budapest, Hungary}
\par\end{center}}
\begin{abstract}
We investigate the ground-state energy of the integrable two dimensional
$O(3)$ sigma model in a magnetic field. By determining a large number
of perturbative coefficients we explore the closest singularities
of the corresponding Borel function. We then confront its median resummation
to the high precision numerical solution of the exact integral equation
and observe that the leading exponentially suppressed contribution
is not related to the asymptotics of the perturbative coefficients.
By analytically expanding the integral equation we calculate the leading
non-perturbative contributions up to fourth order and find complete
agreement. These anomalous terms could be attributed to instantons,
while the asymptotics of the perturbative coefficients seems to be
related to renormalons.
\end{abstract}

\section{Introduction}

Perturbation theory in quantum field theories is typically asymptotic
with factorially growing coefficients. This factorial growth can be
attributed to the proliferation of Feynman diagrams \cite{Hurst:1952zh,Lipatov:1976ny}
related to \emph{instanton}s, or to integrals in specific so called
\emph{renormalon} diagrams, when the loop momenta lie in various IR
and UV domains \cite{Beneke:1998ui}. Instantons correspond to non-trivial
saddle points in the path integral, while renormalons do not have
such a direct semiclassical interpretation. They both lead to non-perturbative
contributions, exponentially suppressed in perturbation theory, which
can typically be extracted from the large-order behaviour of the perturbative
coefficients. The theory which describes this connection is known
as resurgence theory see \cite{Dorigoni:2014hea,Dunne:2015eaa,Aniceto:2018bis,Marino:2012zq}
and references therein. In its strong version it states that all non-perturbative
corrections have their seeds in the perturbative growths. In the present
letter we would like to investigate a counter-example to this behaviour:
the two dimensional $O(3)$ sigma model.

The two-dimensional $O(N)$ sigma models are ideal testing grounds
for non-perturbative physics as they exhibit interesting physical
phenomena including dynamical mass generation and asymptotic freedom,
while they are integrable enabling the exact determination of their
mass gap, scattering matrices, and the ground state energy in a magnetic
field \cite{Zamolodchikov:1977nu,Hasenfratz:1990zz,Hasenfratz:1990ab}.
The $O(3)$ model is exceptional among the other $O(N)$ models as
it is the only one having instantons \cite{Polyakov:1975yp}.

The $O(N)$ models always played a pioneering role in the integrable
investigations: starting from the exact calculation of the scattering
matrix \cite{Zamolodchikov:1977nu} to the exact determination of
the mass gap \cite{Hasenfratz:1990ab,Hasenfratz:1990zz}. This mass
gap was obtained by calculating the groundstate energy in a magnetic
field through perturbation theory and by matching it to the expansion
of the Wiener-Hopf solution of its exact integral equation. Volin's
method allowed to calculate many terms in this expansion whose factorial
asymptotic behaviour revealed the leading singularities on the Borel
plane \cite{Volin:2009wr,Volin:2010cq,Marino:2019eym}. A precision
analysis in the $O(4)$ model showed non-trivial resurgence relations
for the leading non-perturbative behaviour and allowed us to construct
the first few terms in the ambiguity free trans-series, whose median
resummation was in complete agreement with the numerical solution
of the integral equation \cite{Abbott:2020mba,Abbott:2020qnl}. Some
of these analyses was put on analytical grounds in \cite{Bajnok:2021dri}.
There were also analytical developments for large $N$ in \cite{Marino:2021six,DiPietro:2021yxb}.
Recently the authors of \cite{Marino:2021dzn} developed a technique
to calculate the exponentially suppressed corrections systematically
and observed anomalous terms for the $O(3)$ model.

The aim of our present paper is to repeat the analysis we did for
the $O(4)$ model and compare the numerical solution of the integral
equation to the median resummation of the trans-series built from
the asymptotics of the perturbative expansion. As we also observe
these anomalous terms we calculate their expansion beyond the results
of \cite{Marino:2021dzn} and match them to our numerical solution
obtaining complete agreement.

The paper is organized as follows: In the next section we introduce
the integrable description of the ground state energy in the $O(3)$
model, together with Volin's approach, which leads to its perturbative
expansion. Section 3 contains our numerical results. We first solve
Volin's equation numerically and determine the first 336 perturbative
coefficients. Their asymptotic behaviour determines the closest singularities
on the Borel plane, which we can characterize analytically. We then
calculate the lateral Borel resummation of the perturbative series
using the conformal Padé approximation. We subtract the real part
of this result from a high precision solution of the exact integral
equation and investigate the leading exponentially suppressed contributions.
Surprisingly the leading, exponentially small deviation is not related
to the asymptotics of the perturbative coefficients. In section 4
we perform an analytical expansion of the integral equation and determine
the first few non-perturbative terms, which completely agrees with
our numerical results. Finally we conclude in section 5.

\section{O(3) sigma model}

The $O(3)$ sigma model is an integrable two dimensional quantum field
theory, consisting of three particles of mass $m$, whose non-diagonal
scattering matrix is exactly known. By introducing a large enough
magnetic field coupled to one of the $O(3)$ charges one type of particles
can be forced to condense into the vacuum \cite{Hasenfratz:1990zz,Hasenfratz:1990ab}.
The rapidity density of the vacuum condensate satisfies the following
integral equation
\begin{equation}
\chi(B,\theta)-\int_{-B}^{B}\frac{d\theta'}{2\pi}K(\theta-\theta')\chi(B,\theta')=\cosh\theta\quad,\label{eq:integralequation}
\end{equation}
where the kernel is the logarithmic derivative of the condensed particles'
scattering matrix, which takes a very simple form $K(\theta)=2\pi/(\theta^{2}+\pi^{2})$.
The particle density and the energy density can be obtained from the
rapidity density as 
\begin{equation}
\rho(B)=m\int_{-B}^{B}\frac{d\theta}{2\pi}\chi(B,\theta)\quad;\qquad\epsilon(B)=m^{2}\int_{-B}^{B}\frac{d\theta}{2\pi}\cosh\theta\,\chi(B,\theta)\quad.
\end{equation}
The parameter $B$ is related to the magnetic field and a large magnetic
field $h$ corresponds to a large $B$. In order to make contact with
ordinary perturbation theory it is customary to analyze $\epsilon/\rho^{2}$
as the function of the running coupling $\alpha$ which in the $O(3)$
model is defined by 
\begin{equation}
\frac{1}{\alpha}=\ln\frac{m}{\Lambda}+\ln\frac{2\pi\rho}{m}=B+\frac{1}{2}\ln B+\ln4-\frac{1}{2}+\ln\hat{\rho}\quad,\label{eq:runningcoupling}
\end{equation}
where $\Lambda$ is the perturbative scale \cite{Hasenfratz:1990zz},
$\rho=\frac{\sqrt{B}}{\pi}Am\hat{\rho}$ and $A=e^{B+1/2}/4$.

In order to develop a perturbative expansion Volin \cite{Volin:2009wr,Volin:2010cq}
investigated the resolvent $R(\theta)=\int_{-B}^{B}d\theta'\,\frac{\chi(\theta')}{\theta-\theta'}$
and its Laplace transform $\hat{R}(s)=\int_{-i\infty+0}^{i\infty+0}\frac{dz}{2\pi i}e^{sz}R(B+z/2)$,
where $z=2(\theta-B)$. Comparing their two alternative parametrizations
\begin{equation}
R(\theta)=\frac{2A\sqrt{B}}{\theta\sqrt{1-\frac{B^{2}}{\theta^{2}}}}\sum_{n,m=0}^{\infty}\sum_{k=0}^{n+m}\frac{c_{n,m,k}\left(\frac{\theta}{B}\right)^{p(k)}}{B^{m-n}(\theta^{2}-B^{2})^{n}}\left[\log\frac{\theta-B}{\theta+B}\right]^{k}\quad,
\end{equation}
where $p(k)=k\,\mathrm{mod}\,2$ and 
\begin{equation}
\hat{R}(s)=\frac{A}{\sqrt{s}}e^{-s\log(4s/e)}\frac{\Gamma(1+2s)}{\Gamma(\frac{1}{2}+s)}\left(\ensuremath{\frac{1}{s+\frac{1}{2}}+\frac{1}{Bs}\sum_{n,m=0}^{\infty}}\ensuremath{\frac{Q_{n,m}}{B^{m+n}s^{n}}}\right)\quad,
\end{equation}
both the coefficients $c_{n,m,k}$ and $Q_{n,m}$ can be determined
recursively. Once they are known the energy density is $\epsilon/m^{2}=e^{B}\hat{R}(1/2)/(4\pi)$,
while the density is the residue of $R(\theta)$ at infinity. The
first few perturbative orders are 
\begin{equation}
\hat{\rho}=1+\frac{\ln4+\frac{1}{2}\ln B-\frac{3}{4}}{2B}+O(1/B^{2})
\end{equation}
 and 
\begin{equation}
\epsilon(B)=\frac{m^{2}e^{2B+1}}{16\pi}\,\hat{\epsilon}(B)\quad;\quad\hat{\epsilon}(B)=1+\frac{1}{4B}+O(1/B^{2})\quad.
\end{equation}
 Using the definition of $\alpha$ in (\ref{eq:runningcoupling})
allows to calculate the perturbative expansion of 
\begin{equation}
\frac{\epsilon}{\pi\rho^{2}}=\alpha+\frac{\alpha^{2}}{2}+\frac{\alpha^{3}}{2}+\dots=\sum_{n=1}^{\infty}\chi_{n}\alpha^{n}\quad.
\end{equation}

\section{Numerical investigations}

By expanding $R(z)$ in powers of $z$ and performing the Laplace
transform a double series in $s$ and $t=\log(4Bs)$ is obtained containing
$c_{n,m,k}$. On the other hand $\hat{R}(s)$ can be also expanded
similarly and matching the coefficients provides algebraic equations
which can be solved recursively for the $c_{n,m,k}$ and $Q_{n,m}$
coefficients. At each step one can express the $c$ and $Q$ coefficients
in terms of their smaller index counterparts. As the expressions containing
zeta numbers, $\log(2)$ and $\gamma_{E}$ are growing very fast,
it is not possible to go high orders with any reasonable computer
resources. Even taking into account that $\log(2)$ , $\gamma_{E}$
and even $\zeta$ numbers eventually cancel we could not go over 50
coefficients. We then started to determine the coefficients numerically
with a few thousand digits precision. With this approach we could
reach 336 perturbative coefficients in more than a month of computer
time. In the following we analyze these numerical coefficients.

\begin{figure}
\begin{centering}
\includegraphics[width=8cm]{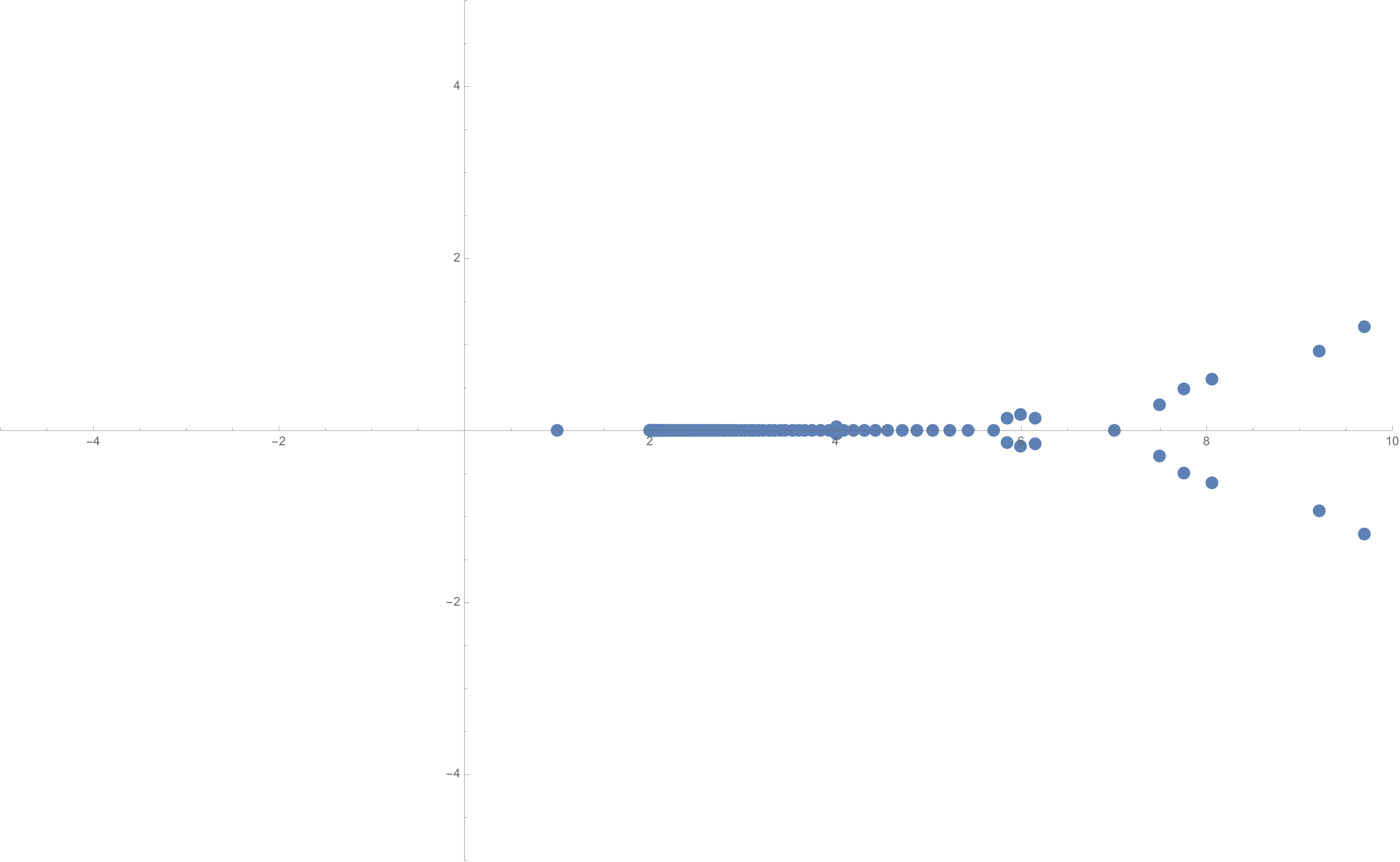}
\par\end{centering}
\caption{Poles of the Padé approximant on the Borel plane.}

\label{figure}
\end{figure}

The coefficients $\chi_{n}$ grow factorially such that 
\begin{equation}
c_{n}=\chi_{n}2^{n-1}/\Gamma(n)
\end{equation}
show constant asymptotics. The poles of the Padé approximant of $B(s)=\sum_{n=1}^{336}c_{n}s^{n-1}$
is demonstrated on Figure \ref{figure}.

The Padé approximant clearly shows a pole singularity at $1$, a cut
starting at $2$, possibly another cut starting at $4$ and may be
more cuts on the positive real lines. It is clear that there is no
singularity on the negative real line in contrast to the $O(N)$ models
for $N>3$, see \cite{Volin:2009wr,Volin:2010cq,Marino:2019eym,Marino:2021dzn}.
Far from the origin our data are not precise enough as we observe
spurious imaginary poles, which change their positions when we vary
the nature or the order of the Padé approximant.

By using high order Richardson transform we can easily extract the
large order behaviour of the $c_{n}$ coefficients
\begin{equation}
c_{n}=\frac{8}{e^{2}}+O(2^{-n})\quad.
\end{equation}
We could see this on the 55th Richardson transform with a 98 digits
precision. We then analysed with similar methods the subleading behaviour
\begin{equation}
(c_{n}-\frac{8}{e^{2}})\frac{2^{n-1}}{n}=a_{0}+\frac{a_{1}}{n}+\frac{a_{2}}{n(n-1)}+\dots\frac{a_{k}}{n(n-1)\dots(n-k+1)}+\dots
\end{equation}
and found for 50 digits that 
\begin{equation}
a_{0}=-\frac{32}{\pi e^{4}}\quad;\qquad\frac{a_{1}}{a_{0}}=-\frac{3}{2}\quad;\qquad\frac{a_{2}}{a_{0}}=\frac{1}{8}\quad.
\end{equation}
Similarly we found

\begin{equation}
\frac{a_{3}}{a_{0}}=-\frac{127}{16}+6\zeta(3)\quad;\quad\frac{a_{4}}{a_{0}}=-\frac{35087}{384}+\frac{137\zeta(3)}{2}
\end{equation}
\begin{equation}
\frac{a_{5}}{a_{0}}=\frac{1005\zeta(3)}{2}+\frac{405\zeta(5)}{2}-\frac{226801}{256}
\end{equation}
with decreasing precisions. We also calculated the next 20 coefficients
numerically and from the numerics we could observe that they behave
as 
\begin{equation}
a_{n}/a_{0}=\Gamma(n+2)(b_{0}+\frac{b_{1}}{n}+\dots)\quad,
\end{equation}
but we could calculate the $b_{n}$ coefficients for $n=0,1,$ only
with a few digits precision. $b_{0}$ seems to be $\frac{32}{\pi e^{4}}$
with 3 digits while $b_{1}/b_{0}\sim5/2$ with two digits.

We then decided to compare the lateral Borel resummations of our perturbative
answer to the numerical solution of the TBA equation. In the first
step we elaborated on the numerical evaluation of the inverse Borel
transform. If we had just used the Padé approximant of the perturbative
series we would have faced the complex singularities (see Figure \ref{figure})
on any ray of integrations. So we decided to use the conformal mapping
$s=4w/(1+w)^{2}$ to rewrite the series on the Borel plane in powers
of $w$. Technically we worked with 
\begin{equation}
f(\alpha)=\frac{\epsilon}{\pi\rho^{2}}=\sum_{n=1}^{336}\chi_{n}\alpha^{n}\quad;\qquad B(s)=\sum_{n=0}^{335}c_{n+1}s^{n}\quad.
\end{equation}
By using the conformal mapping we can calculate the $d_{n}$ coefficients
from 
\begin{equation}
B(s)=\sum_{n}d_{n}w(s)^{n}\quad;\qquad w(s)=\frac{1-\sqrt{1-s}}{1+\sqrt{1-s}}
\end{equation}
by matching their perturbative $s$-expansions. We then can integrate
\begin{equation}
f_{+}(\alpha)=2\int_{0}^{e^{i/2}\infty}\sum_{n}d_{n}w(s)^{n}e^{-2s/\alpha}ds\quad.\label{eq:fplus}
\end{equation}
We could even make a Padé approximation for $B(w)$ but it did not
really change the result of the integral which we calculated at least
for 50 digits.

We then solved the integral equation with a very high precision similar
to \cite{Abbott:2020qnl}. This was done by expanding $\chi$ in a
Chebisev basis on the interval $[-B,B]$. Due to the rational nature
of the kernel we could calculate its matrix elements analytically
and by inverting the matrix equation we calculated the representative
of $\chi$. We managed to get $f_{\mathrm{TBA}}=\frac{\epsilon}{\pi\rho^{2}}$
as a function of $\alpha$ with 70 digits precision (in the worst
case). We then compared $f_{\mathrm{TBA}}$ to $f_{+}$ separately
for the imaginary and for the real part. The imaginary part started
as 
\begin{equation}
\Im m(f_{+})=\frac{16\pi}{e^{2}}e^{-2/\alpha}+\frac{e^{-4/\alpha}}{\alpha}\sum_{n=0}\tilde{a}_{n}\alpha^{n}\quad,
\end{equation}
where we found that $\tilde{a}_{n}=\pi a_{n}4^{2-n}$ for $n=0,\dots,5$
with decreasing but at least 5 digits precision. Median resummation
similar to \cite{Abbott:2020qnl} would suggest the form of the real
deviation

\begin{equation}
f_{\mathrm{TBA}}-\Re e(f_{+})\propto e^{-8/\alpha}(4b_{0}+b_{1}\alpha+\dots)\quad,
\end{equation}
however we observed a deviation of a much earlier order of the form
\begin{equation}
f_{\mathrm{TBA}}-\Re e(f_{+})=e^{-2/\alpha}A_{0}(2/\alpha+A_{1}+A_{2}\log\alpha+\alpha A_{3}+O(\alpha^{2}))\quad.
\end{equation}
Originally we tried to fit a polynomial in $\alpha$ and failed to
get a reasonable precision. In the recent paper \cite{Marino:2021dzn}
the authors calculated the leading coefficients to be 
\begin{equation}
A_{0}=\frac{32}{e^{2}}\quad;\quad A_{1}=3-\gamma_{E}-5\log2\quad;\quad A_{2}=1\quad.
\end{equation}
By subtracting these coefficients we could fit $A_{3}=\frac{1}{2}$
with 5 digits and observed that there are no logarithmic correction
at this order.

In order to test this prediction we expanded the integral equation
one order higher than \cite{Marino:2021dzn} and showed that indeed
the only correction at this order is $A_{3}=\frac{1}{2}$. We summarize
the details of the calculation in the next section.

\section{Leading exponential corrections from the integral equation}

The integral equation can be solved using the Wiener-Hopf technique.
Here we summarize our results, while the details of the calculation
will be given elsewhere. The idea of the calculation is to use Fourier
transformation. If the integral went for the whole line we could easily
calculate the Fourier transform of the kernel $\tilde{K}(\omega)=\int d\theta K(\theta)e^{i\omega\theta}$
and act with the inverse of $1-\tilde{K}(\omega)$ on the source 
\begin{equation}
g(\omega)=\frac{e^{B}}{2i}\left(\frac{e^{iB\omega}}{\omega-i}-\frac{e^{-iB\omega}}{\omega+i}\right)+\frac{e^{-B}}{2i}\left(\frac{e^{iB\omega}}{\omega+i}-\frac{e^{-iB\omega}}{\omega-i}\right)\quad.
\end{equation}
Since $\chi(B,\theta)$ is non-vanishing only on the interval $[-B,B]$
the integral equation is not satisfied for $\vert\theta\vert>B$.
In the Wiener-Hopf technique we introduce the missing function and
determine it by separating the equations in Fourier space into terms
analytic on the lower and upper half planes, see \cite{Marino:2021dzn}
for the details in the present context. This requires to find the
factorization 
\begin{equation}
(1-\tilde{K}(\omega))^{-1}=G_{+}(\omega)G_{-}(\omega)\quad,
\end{equation}
where $G_{+}(\omega)$, $G_{-}(\omega)=G_{+}(-\omega)$ are analytic
on the upper/lower half plane. We also need integral operators $f(\omega)_{\pm}=\pm\frac{i}{2\pi}\int_{-\infty}^{\infty}\frac{f(\omega')d\omega'}{\omega-\omega'\pm i0}$
to project on the corresponding components. This integral contour
then can be deformed to surround the singularities of $G_{-}(\omega)$
and $G_{-}(\omega)/G_{+}(\omega)$, which are on the positive imaginary
line. The speciality of the $O(3)$ model is that 
\begin{equation}
G_{-}(\omega)=\frac{(2e)^{\frac{i\omega}{2}}}{\sqrt{\pi}}(i\omega)^{\frac{1}{2}i(i-\omega)}\Gamma\left(1+\frac{i\omega}{2}\right)
\end{equation}
is nonvanishing at $i$: $G_{-}(i)=1/\sqrt{2e}$ and the pole $1/(\omega-i)$
will contribute from the source term. This term will appear in the
unknown function and also in the Fourier transform of $\chi$ which
should be taken at $i$ to get the ground state energy.

\begin{figure}
\begin{centering}
\includegraphics[width=8cm]{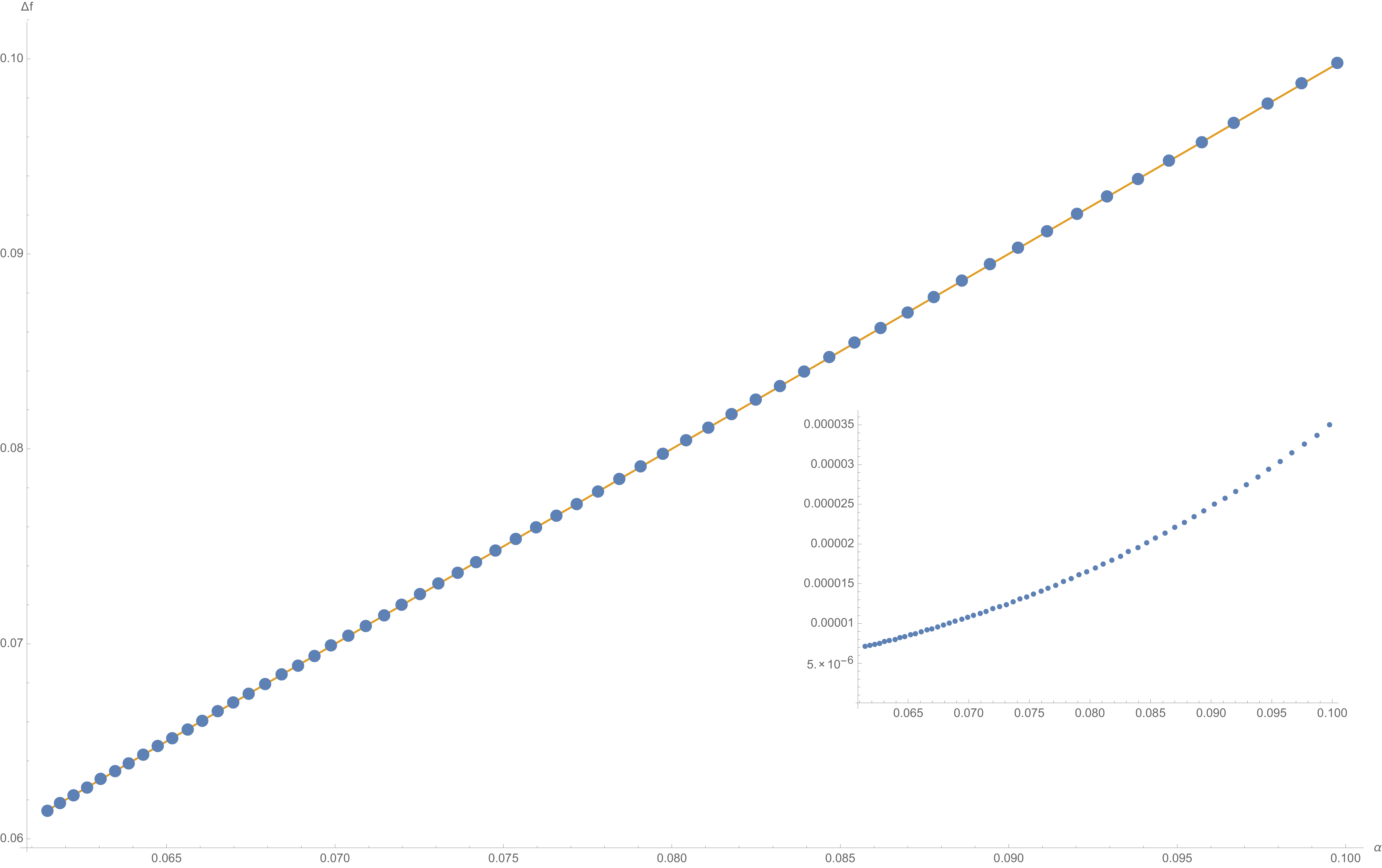}
\par\end{centering}
\caption{Comparison of the numerically determined $f_{\mathrm{TBA}}-\Re e(f_{+})$
to the analytically calculated new term $\alpha/2$. In particular
$\Delta f=e^{2/\alpha}e^{2}/16(f_{\mathrm{TBA}}-\Re e(f_{+}))-2(2/\alpha+3-\gamma_{E}-5\log2+\log\alpha)$
is plotted against $\alpha$ with dots, while the newly calculated
$\alpha$ term with a continous line. The inset demonstrates the difference
between the numerics and the new analytical term.}

\label{fig:alpha}
\end{figure}

After redefining some functions we need to solve the integral equations
\begin{equation}
q^{(\pm)}(x)+\frac{1}{\pi}\int_{{\cal C}_{+}}\frac{e^{-y}\beta(\frac{y}{4B})q^{(\pm)}(y)}{x+y}\,dy=\frac{1}{1\pm\frac{x}{2B}}\quad,
\end{equation}
where $\beta(x)=i(G_{-}(2ix+0)-G_{-}(2ix-0))/2G_{+}(2ix)$ and the
${\cal C}_{+}$ integration goes from zero to infinity, avoiding the
poles of the integrand at $y=4B,8B,\dots$ from the left. This contour
is equivalent to the integration path chosen for the perturbative
part in (\ref{eq:fplus}). From the solutions $q^{(\pm)}$ the leading
exponential correction of $\hat{\epsilon}=\hat{\epsilon}_{\mathrm{pert}}+e^{-2B}\delta\hat{\epsilon}$
can be calculated via 
\begin{equation}
u^{(\pm)}=\frac{1}{\pi}\int_{{\cal C}_{+}}\frac{e^{-y}\beta(\frac{y}{4B})q^{(\pm)}(y)}{2B\mp y}\,dy
\end{equation}
as 
\begin{equation}
\begin{split}\delta\hat{\epsilon} & =\frac{-i\pi}{e}+\frac{2}{e}[2B-\ln2-\gamma_{E}+1-u^{(+)}-u^{(-)}]\\
 & =\frac{-i\pi}{e}+\frac{2}{e}[2B-\ln2-\gamma_{E}+1-\frac{1}{4B}+O(1/B^{2})]\quad.
\end{split}
\label{eq:epsilon}
\end{equation}
Although the density $\rho$ is related to the Fourier transform of
$\chi$ at the origin, it is simpler to extract it from the solution
of an integral equation similar to (\ref{eq:integralequation}) but
with a source term being $1$ instead of $\cosh\theta$. Repeating
the calculation for $\hat{\rho}=\hat{\rho}_{\mathrm{pert}}+e^{-2B}\delta\hat{\rho}$
gives 
\begin{equation}
\delta\hat{\rho}=-\frac{1}{e}\left\{ 1+\frac{\ln4+\frac{1}{2}\ln B+\frac{3}{4}}{2B}+O(1/B^{2})\right\} \quad.\label{eq:rho}
\end{equation}
Due to this term the running coupling (\ref{eq:runningcoupling})
acquires a non-perturbative correction, too. Thus the sought for quantity
$\epsilon/\rho^{2}$ receives corrections from three sources. It gets
corrections directly from (\ref{eq:epsilon}) and from (\ref{eq:rho})
and indirectly from the running coupling (\ref{eq:runningcoupling})
leading finally to 
\begin{equation}
\frac{\epsilon}{\pi\rho^{2}}=\alpha+\frac{\alpha^{2}}{2}+\dots+e^{-\frac{2}{\alpha}}\frac{32}{e^{2}}\left(-\frac{i\pi}{2}+\frac{2}{\alpha}+3-\gamma_{E}-5\log2+\log\alpha+\frac{\alpha}{2}+\dots\right)+O(e^{-\frac{4}{\alpha}})\quad.
\end{equation}
Here the asymptotic perturbative series is understood as Borel resummed
by (\ref{eq:fplus}), thus the imaginary non-perturbative part from
the integral nicely cancels with the similar term coming from the
pole term. The $\alpha/2$ term in the bracket is our new result compared
to \cite{Marino:2021dzn}. We compare this newly calculated $\alpha/2$
term to the numerically determined $f_{\mathrm{TBA}}-\Re e(f_{+})$
on Figure \ref{fig:alpha} and find complete agreement. In the inset
we plot the difference of these two, which signals that the next order
terms are much smaller.

\section{Conclusion}

In this letter we investigated the groundstate energy of the $O(3)$
sigma model both numerically and analytically. On the numerical side
we calculated the first 336 perturbative coefficients with very high
precisions. Their asymptotic behaviour allowed us to map and characterize
the closest singularities on the Borel plane: a single pole and a
cut signaling imaginary ambiguities of the form $\tilde{a}e^{-2/\alpha}$
and $e^{-4/\alpha}/\alpha(\tilde{a}_{0}+\tilde{a}_{1}\alpha+\dots.$).
We identified the appearing coefficients $\tilde{a}$ with high precision
in terms of $16/e^{2}$ and odd zeta functions. We even investigated
their asymptotics which contributes via the median resummation to
the real part of the groundstate energy of order $e^{-8/\alpha}$,
see \cite{Abbott:2020qnl} for a similar analysis in the $O(4)$ model.
We then compared the median resummation of the perturbative and non-perturbative
terms originating from the asymptotics of the perturbative series
to the high precision solution of the integral equations and observed
an anomalous $e^{-2/\alpha}$ deviation, which cannot be attributed
in any way to the perturbative coefficients. We then, following \cite{Marino:2021dzn},
expanded analytically the integral equation and matched the results
against the observed deviation and found complete agreement.

Our analytical calculation differs technically, but not conceptually
from the calculation in \cite{Marino:2021dzn} since we calculated
$\epsilon/\rho^{2}$ as the function of $\rho(B)$, while they calculated
the free energy as the function of the magnetic field. These quantities
are related by a Legendre transformation. Furthermore, our calculation
goes one order beyond the result of \cite{Marino:2021dzn}.

These results strongly indicate that the anomalous non-perturbative
terms are not related to the asymptotics of the perturbative series.
This is exceptional for the $O(3)$ model and does not appear for
any higher $N$ \cite{Abbott:2020mba,Abbott:2020qnl,Marino:2021dzn,Marino:2021six,DiPietro:2021yxb}.
Since the $O(3)$ model is the only one with instantons, we believe
that the anomalous terms are related to the instantons, while the
regular terms are related to renormalons. Indeed such terms for other
$O(N)$ models in the large $N$ limit were matched to renormalon
diagrams in perturbation theory \cite{Marino:2021six,DiPietro:2021yxb}.
It would be extremely interesting to match the observed anomalous
non-perturbative behaviour with explicit instanton calculations.

\subsection*{Acknowledgements}

Our work was supported by ELKH, with infrastructure provided by the
Hungarian Academy of Sciences. This work was supported by NKFIH grant
K134946.

\bibliographystyle{elsarticle-num}
\bibliography{borel}

\begin{thebibliography}{10}
\expandafter\ifx\csname url\endcsname\relax
  \def\url#1{\texttt{#1}}\fi
\expandafter\ifx\csname urlprefix\endcsname\relax\def\urlprefix{URL }\fi
\expandafter\ifx\csname href\endcsname\relax
  \def\href#1#2{#2} \def\path#1{#1}\fi

\bibitem{Hurst:1952zh}
C.~A. Hurst, {The Enumeration of Graphs in the Feynman-Dyson Technique}, Proc.
  Roy. Soc. Lond. A 214 (1952) 44.
\newblock \href {https://doi.org/10.1098/rspa.1952.0149}
  {\path{doi:10.1098/rspa.1952.0149}}.

\bibitem{Lipatov:1976ny}
L.~Lipatov, {Divergence of the Perturbation Theory Series and the
  Quasiclassical Theory}, Sov. Phys. JETP 45 (1977) 216--223.

\bibitem{Beneke:1998ui}
M.~Beneke, {Renormalons}, Phys. Rept. 317 (1999) 1--142.
\newblock \href {http://arxiv.org/abs/hep-ph/9807443}
  {\path{arXiv:hep-ph/9807443}}, \href
  {https://doi.org/10.1016/S0370-1573(98)00130-6}
  {\path{doi:10.1016/S0370-1573(98)00130-6}}.

\bibitem{Dorigoni:2014hea}
D.~Dorigoni, {An Introduction to Resurgence, Trans-Series and Alien Calculus},
  Annals Phys. 409 (2019) 167914.
\newblock \href {http://arxiv.org/abs/1411.3585} {\path{arXiv:1411.3585}},
  \href {https://doi.org/10.1016/j.aop.2019.167914}
  {\path{doi:10.1016/j.aop.2019.167914}}.

\bibitem{Dunne:2015eaa}
G.~V. Dunne, M.~\"Unsal, {What is QFT? Resurgent trans-series, Lefschetz
  thimbles, and new exact saddles}, PoS LATTICE2015 (2016) 010.
\newblock \href {http://arxiv.org/abs/1511.05977} {\path{arXiv:1511.05977}},
  \href {https://doi.org/10.22323/1.251.0010} {\path{doi:10.22323/1.251.0010}}.

\bibitem{Aniceto:2018bis}
I.~Aniceto, G.~Basar, R.~Schiappa, {A Primer on Resurgent Transseries and Their
  Asymptotics}, Phys. Rept. 809 (2019) 1--135.
\newblock \href {http://arxiv.org/abs/1802.10441} {\path{arXiv:1802.10441}},
  \href {https://doi.org/10.1016/j.physrep.2019.02.003}
  {\path{doi:10.1016/j.physrep.2019.02.003}}.

\bibitem{Marino:2012zq}
M.~Mari\~no, {Lectures on non-perturbative effects in large $N$ gauge theories,
  matrix models and strings}, Fortsch. Phys. 62 (2014) 455--540.
\newblock \href {http://arxiv.org/abs/1206.6272} {\path{arXiv:1206.6272}},
  \href {https://doi.org/10.1002/prop.201400005}
  {\path{doi:10.1002/prop.201400005}}.

\bibitem{Zamolodchikov:1977nu}
A.~B. Zamolodchikov, A.~B. Zamolodchikov, {Relativistic Factorized S Matrix in
  Two-Dimensions Having O(N) Isotopic Symmetry}, JETP Lett. 26 (1977) 457.
\newblock \href {https://doi.org/10.1016/0550-3213(78)90239-0}
  {\path{doi:10.1016/0550-3213(78)90239-0}}.

\bibitem{Hasenfratz:1990zz}
P.~Hasenfratz, M.~Maggiore, F.~Niedermayer, {The Exact mass gap of the O(3) and
  O(4) nonlinear sigma models in d = 2}, Phys. Lett. B 245 (1990) 522--528.
\newblock \href {https://doi.org/10.1016/0370-2693(90)90685-Y}
  {\path{doi:10.1016/0370-2693(90)90685-Y}}.

\bibitem{Hasenfratz:1990ab}
P.~Hasenfratz, F.~Niedermayer, {The Exact mass gap of the O(N) sigma model for
  arbitrary $ N \geq 3$ in d = 2}, Phys. Lett. B 245 (1990) 529--532.
\newblock \href {https://doi.org/10.1016/0370-2693(90)90686-Z}
  {\path{doi:10.1016/0370-2693(90)90686-Z}}.

\bibitem{Polyakov:1975yp}
A.~M. Polyakov, A.~A. Belavin, {Metastable States of Two-Dimensional Isotropic
  Ferromagnets}, JETP Lett. 22 (1975) 245--248.

\bibitem{Volin:2009wr}
D.~Volin, {From the mass gap in O(N) to the non-Borel-summability in O(3) and
  O(4) sigma-models}, Phys. Rev. D 81 (2010) 105008.
\newblock \href {http://arxiv.org/abs/0904.2744} {\path{arXiv:0904.2744}},
  \href {https://doi.org/10.1103/PhysRevD.81.105008}
  {\path{doi:10.1103/PhysRevD.81.105008}}.

\bibitem{Volin:2010cq}
D.~Volin, {Quantum integrability and functional equations: Applications to the
  spectral problem of AdS/CFT and two-dimensional sigma models}, Ph.D. thesis
  (2009).
\newblock \href {http://arxiv.org/abs/1003.4725} {\path{arXiv:1003.4725}},
  \href {https://doi.org/10.1088/1751-8113/44/12/124003}
  {\path{doi:10.1088/1751-8113/44/12/124003}}.

\bibitem{Marino:2019eym}
M.~Mari\~no, T.~Reis, {Renormalons in integrable field theories}, JHEP 04
  (2020) 160.
\newblock \href {http://arxiv.org/abs/1909.12134} {\path{arXiv:1909.12134}},
  \href {https://doi.org/10.1007/JHEP04(2020)160}
  {\path{doi:10.1007/JHEP04(2020)160}}.

\bibitem{Abbott:2020mba}
M.~C. Abbott, Z.~Bajnok, J.~Balog, A.~Heged{\H u}s, {From perturbative to
  non-perturbative in the O(4) sigma model}, Phys. Lett. B 818 (2021) 136369.
\newblock \href {http://arxiv.org/abs/2011.09897} {\path{arXiv:2011.09897}},
  \href {https://doi.org/10.1016/j.physletb.2021.136369}
  {\path{doi:10.1016/j.physletb.2021.136369}}.

\bibitem{Abbott:2020qnl}
M.~C. Abbott, Z.~Bajnok, J.~Balog, A.~Heged{\H u}s, S.~Sadeghian, {Resurgence
  in the O(4) sigma model}, JHEP 05 (2021) 253.
\newblock \href {http://arxiv.org/abs/2011.12254} {\path{arXiv:2011.12254}},
  \href {https://doi.org/10.1007/JHEP05(2021)253}
  {\path{doi:10.1007/JHEP05(2021)253}}.

\bibitem{Bajnok:2021dri}
Z.~Bajnok, J.~Balog, I.~Vona, {Analytic resurgence in the O(4) model}, 
\newblock \href {http://arxiv.org/abs/2111.15390} {\path{arXiv:2111.15390}}.

\bibitem{Marino:2021six}
M.~Marino, R.~Miravitllas, T.~Reis, {Testing the Bethe ansatz with large N
  renormalons} (2 2021).
\newblock \href {http://arxiv.org/abs/2102.03078} {\path{arXiv:2102.03078}},
  \href {https://doi.org/10.1140/epjs/s11734-021-00252-4}
  {\path{doi:10.1140/epjs/s11734-021-00252-4}}.

\bibitem{DiPietro:2021yxb}
L.~Di~Pietro, M.~Mari\~no, G.~Sberveglieri, M.~Serone, {Resurgence and 1/N
  Expansion in Integrable Field Theories}, JHEP 10 (2021) 166.
\newblock \href {http://arxiv.org/abs/2108.02647} {\path{arXiv:2108.02647}},
  \href {https://doi.org/10.1007/JHEP10(2021)166}
  {\path{doi:10.1007/JHEP10(2021)166}}.

\bibitem{Marino:2021dzn}
M.~Marino, R.~Miravitllas, T.~Reis, {New renormalons from analytic
  trans-series}, 
\newblock \href {http://arxiv.org/abs/2111.11951} {\path{arXiv:2111.11951}}.

\end{thebibliography}

\end{document}